\documentclass[11pt]{article}   	
\usepackage{geometry}                		
 \usepackage{graphicx}			 	
\usepackage{amssymb}
\usepackage{amsmath}
\DeclareMathOperator*{\argmin}{argmin} 

\newcommand{\Reals}{{\mathrm{I}\!\mathrm{R}}}
\newcommand{\Bbeta}{\boldsymbol{\beta}}
\newcommand{\Bbetahat}{\hat{\boldsymbol{\beta}}}

\newcommand{\Xvec}{\vec{\boldsymbol{X}}}
\newcommand{\xvec}{\vec{\boldsymbol{x}}}
\renewcommand{\P}{\boldsymbol{P}}
\newcommand{\PX}{\boldsymbol{P}_{\!\!\Xvec}}
\newcommand{\PYX}{\boldsymbol{P}_{\!\!Y,\Xvec}}
\newcommand{\PYgX}{\boldsymbol{P}_{\!\!Y|\Xvec}}
\newcommand{\Phat}{\hat{\boldsymbol{P}}_{\!n}}
\newcommand{\E}{\boldsymbol{E}}
\newcommand{\AV}{\boldsymbol{AV}}
\newcommand{\AVhat}{\widehat{\boldsymbol{AV}}}

\newcommand{\logit}{\mathrm{logit}}

\title{Assumption Lean Regression}
\author{Richard Berk, Andreas Buja, Lawrence Brown, Edward George  
\\ Arun Kumar Kuchibhotla, Weijie Su, and Linda Zhao \\ Department of Statistics \\
University of Pennsylvania}
\begin{document}
\maketitle

\begin{abstract}
It is well known that models used in conventional regression analysis are
commonly misspecified. A standard response  is little more than a
shrug. Data analysts invoke Box's maxim that all models are wrong and 
then proceed as if the results are useful nevertheless. In this paper, we provide 
an alternative. Regression models are treated explicitly as approximations
of a true response surface that can have a number of desirable statistical 
properties, including estimates that are asymptotically unbiased. Valid statistical 
inference follows. We generalize the formulation to include regression functionals,
which broadens substantially the range of potential applications. An empirical
application is provided to illustrate the paper's key concepts. 
\end{abstract}

\section{Introduction} \label{sec:intro}

It is old news that models are approximations and that regression
analyses of real data commonly employ models that are misspecified in
various ways.  Conventional approaches are laden with assumptions
that are questionable, many of which are effectively untestable (Box,
1976, Leamer, 1878; Rubin, 1986; Cox, 1995; Berk, 2003; Freedman,
2004; 2009).  This note discusses some implications of an ``assumption
lean'' reinterpretation of regression.  In this reinterpretation, one
requires only that the observations are i.i.d., realized at random
according to a joint probability distribution of the regressor and
response variables.  If no model assumptions are made, then the
parameters of fitted models need to be interpreted as statistical
functionals, here called ``regression functionals.''

For ease and clarity of exposition we begin with linear regression.
Later we turn to other types of regression and show how the lessons
from linear regression carry forward to the generalized linear model
and even more broadly.  We draw heavily on two papers by Buja et
al.~(2016a;b), a portion of which draws on early insights of Halbert
White~(1980a). 

\section{The Parent Joint Probability Distribution} \label{sec:parent}

For observational data, suppose there is a set of
quantitative random variables that have a joint distribution $\P$,
also called the ``population,'' that characterizes regressor variables
$X_{1}, \dots, X_{p}$ and a response variable $Y$.  The distinction
between regressors and the response is determined by the data analyst
based on subject matter interest.  These designations do not imply any
causal mechanisms and or any particular generative models for
$\P$.  Unlike in linear models theory, the regressor variables are
not interpreted as fixed: they are as random as the response and
will treated as such.  

We collect the regressor variables in a $(p\!+\!1) \times 1$ column random vector
$\Xvec = (1,X_{1} \dots, X_{p})'$ with a leading 1 to accommodate an
intercept in linear models.  We write $\P = \PYX$ for the joint
probability distribution, $\PYgX$ for the conditional distribution of
$Y$ given $\Xvec$, and $\PX$ for the marginal distribution of~$\Xvec$.
The only assumption made is that the data are realized
i.i.d.~from $\P$.  The separation of the random variables into
regressors and a response implies that there is interest in $\PYgX$.
Hence, some form of regression analysis is applied.  Yet, because the
regressors are random variables, their marginal distribution $\PX$
cannot be ignored.

\section{Estimation Targets} \label{sec:targets}

As a feature of $\P$ or, more precisely, of $\PYgX$, there is a ``true
response surface'' denoted by $\mu(\Xvec)$.  Most often, $\mu(\Xvec)$
is the conditional expectation of $Y$ given $\Xvec$,
$\mu(\Xvec) = \E[Y|\Xvec]$, but there are other possibilities,
depending on the context. For example, $\mu(\Xvec)$ might be chosen
to be the conditional median or some other conditional quantile of $Y$
given $\Xvec$.  The true response surface is a common estimation
target for conventional regression in which a data analyst assumes a
specific parametric form.  We will \textit{not} proceed in this manner
and will not make assumptions about what form $\PYgX$ actually takes.
Yet, we will make use of standard ordinary least squares (OLS) fitting
of linear equations.  This approach reflects data analytic situations in which
either deviations from linearity in $\mu(\Xvec)$ may be difficult to
detect with diagnostics, or in which the fitted linear formula is known
to be a deficient approximation to $\mu(\Xvec)$, and yet, OLS is employed
because of underlying substantive theories, measurement
requirements, or considerations of interpretability.

Fitting a linear function $l(\Xvec) = \Bbeta' \Xvec$ to $Y$ with OLS
can be achieved mathematically at the population $\P$ without assuming
that the response surface $\mu(\Xvec)$ is linear in~$\Xvec$:
\begin{equation} \label{eq:population-OLS}
\Bbeta (\P) = \argmin_{\Bbeta \in \Reals^{p+1}} \E[(Y - \Bbeta' \Xvec)^{2}] .
\end{equation}
The vector $\Bbeta = \Bbeta(\P)$ is the ``population OLS solution''
and contains the ``population coefficients.''  Notationally, when we
write $\Bbeta$, it is understood to be $\Bbeta(\P)$.  Similar to
finite datasets, the OLS solution for the population can be obtained
by solving a population version of the normal equations, resulting in
\begin{equation} \label{eq:population-tripleX}
\Bbeta(\P) = \E[\Xvec \Xvec']^{-1} \E[ \Xvec Y] .
\end{equation}
Thus, one obtains the best linear approximation in the OLS sense to $Y$
as well as to~$\mu(\Xvec)$.  As such, it should be useful without
(unrealistically) assuming that $\mu(\Xvec)$ is identical to
$\Bbeta'\Xvec$.  

We have worked so far with a distribution/population $\P$, not data.  
We have, therefor, defined a target of estimation: $\Bbeta(\P)$ obtained from
\eqref{eq:population-OLS} and \eqref{eq:population-tripleX} is the
estimand of empirical OLS estimates $\hat{\Bbeta}$ obtained from data.
This estimand is well-defined as long as the joint distribution $\P$
has second moments and the regressor distribution $\PX$ is not
perfectly collinear. That is, the second moment matrix
$\E[\Xvec\Xvec']$ is full rank.  No other assumptions are needed. In
particular, there are no assumptions of linearity of~$\mu(\Xvec)$,
homoskedasticity, or Gaussianity.  This constitutes the
``assumption lean'' or ``model robust'' framework.

A foundational question is why one should settle for the best linear
{\em approximation} to the truth.  Indeed, those who insist that models 
must always be ``correctly specified'' will be unreceptive. They may 
insist that models should be revised until diagnostics and goodness of 
fit tests no longer detect deficiencies.  One may then legitimately proceed 
as if the model is correct. 

Such thinking warrants careful scrutiny. Data analysis with a given, fixed sample 
size requires decisions about how to balance the desire for good models against 
the costs of data dredging. ``Improving'' models by searching regressors, 
trying out transformations of all variables, inventing new regressors from 
existing ones, applying algorithms and interactive experiments, and undertaking 
assessments with diagnostic tests and plots can each invalidate subsequent 
statistical inference. The result often is models that not only fit the data well, 
but fit them too well (Hong et al. 2017).  

Research is underway to provide valid  post-selection inference (e.g., 
Berk et al.~2013, Lee et al.~2016), which is an important special case. 
But the proposed procedures address solely regressor selection 
and typically make strong assumptions. With these significant caveats, 
asymptotically valid post-selection inference under misspecification 
has substantial promise (Bachoc et al.~2016, Kuchibhotla et al.~2018), 
but there is not yet much to help the data analyst.

Beyond the costs of data dredging, there can be substantive reasons
for curtailing ``model improvement.'' Some variables may express
phenomena in ``natural'' or ``conventional'' units that should not
be transformed even if model fit is improved. A substantive theory may require 
a particular model that does not fit the data well. Identifying important 
variables may be the primary concern, making quality of the fit less important. 
Predictors prescribed by subject-matter theory or past research may be 
unavailable so that the model specified is the best that can be done.
In short, one must consider ways in which valid statistical inference can
be undertaken with models acknowledged to be approximations.

Note that we are {\bf\em not} making an argument for
discarding model diagnostics. It is always important to learn
what one can from the data, including model deficiencies that
properly circumscribe conclusions being drawn. But there
can be serious risks trying impose remedies that really 
are not.

We also are not simply restating Box's maxim that models are 
always ``wrong'' in some ways but can useful despite their deficiencies. 
Acknowledging models as approximations is one thing. Understanding the
consequences is another.  What follows, therefore, is a discussion of some 
of these consequences and an argument in favor of assumption lean
inference employing model robust standard errors, such as those
obtained from sandwich estimators or the $x$-$y$ bootstrap.

\section{A Population Decomposition of the Conditional Distribution of $Y$} 
\label{sec:decomp}

\begin{figure}[t]
\begin{center}
   \includegraphics[width=4.8in]{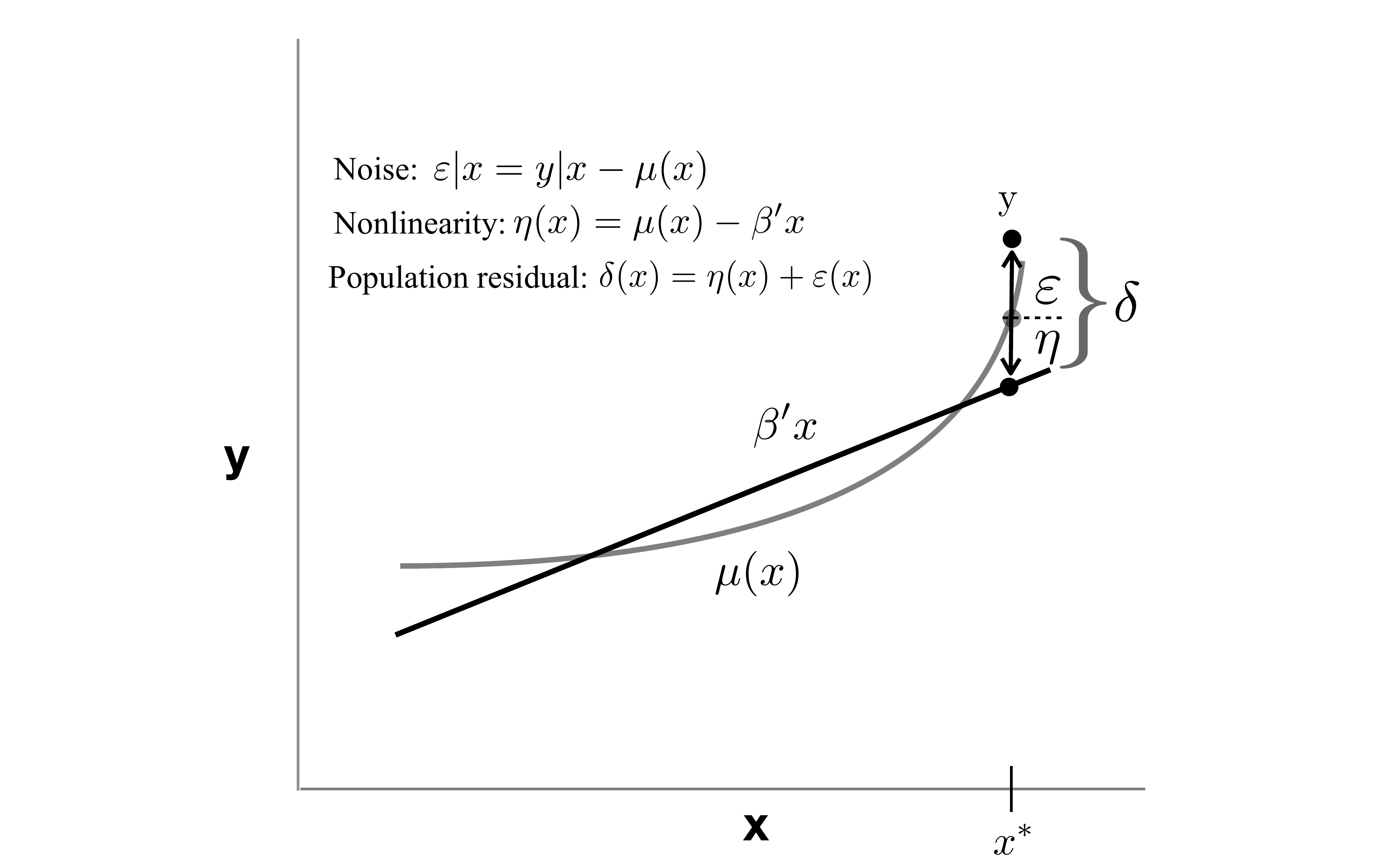} 
   \caption{A Population Decomposition of $Y|X$ Using the Best Linear Approximation}
\label{fig:decomp}
\end{center}
\end{figure}

A first step in understanding the statistical properties of the best
linear approximation is to consider carefully the potential
disparities in the population between $\mu(\Xvec)$ and $\Bbeta'\Xvec$.
Figure~\ref{fig:decomp} provides a visual representation.  There is
for the moment a response variable $Y$ and a single regressor~$X$.

The curved line shows the true response surface $\mu(x)$.  The
straight line shows the best linear approximation
$\beta_0 + \beta_1 x$.  Both are features of the joint probability
distribution, not a realized dataset.

The figure shows a regressor value $x^*$ drawn from $\PX$
and a response value $y$ drawn from $\P_{\!\!Y|X=x^*}$.  The disparity
between $y$ and the fitted value from the best linear approximation is
denoted as $\delta = y - (\beta_0 + \beta_1 x^*)$ and will be called
the ``population residual.''  The value of $\delta$ at $x^*$ can be
decomposed into two components:
\begin{itemize}
\item The first component results from the disparity between the true
  response surface, $\mu(x^*)$, and the approximation
  $\beta_0 +\beta_1 x^*$.  We denote this disparity by
  $\eta = \eta(x^*)$ and call it ``the nonlinearity.''  Because
  $\beta_0 + \beta_1 x^*$ is an approximation, disparities should be
  expected.  They are the result of mean function misspecification.
  As a function of the random variable $X$, the nonlinearity 
  $\eta(X)$ is a random variable as well.
\item The second component of $\delta$ at $x^*$, denoted by
  $\varepsilon$, is random variation around the true conditional mean
  $\mu(x^*)$.  We prefer for such variation the term ``noise'' over
  ``error.''  Sometimes it is called ``irreducible variation'' because
  it exists even if the true response surface is known.
\end{itemize}
The components defined here and shown in Figure~\ref{fig:decomp}
generalize to regression with arbitrary numbers of regressors, in
which case we write $\delta = Y - \Bbeta' \Xvec$,
$\eta = \mu(\Xvec) - \Bbeta' \Xvec$ and
$\varepsilon = Y - \mu(\Xvec)$.  These random variables have
properties with important implications.  Foremost, the population
residual, the nonlinearity and the noise are all
``population-orthogonal'' to the regressors:
\begin{equation} \label{eq:orthogonality}
\E ( X_j \, \delta ) = \E ( X_j \, \eta(\Xvec) ) = \E ( X_j \, \varepsilon ) = 0 ,
\end{equation}
where following the convention introduced in Section~\ref{sec:parent},
the index $j=0$ indicates the intercept, $X_0 = 1$, and $j=1,\ldots,p$
indicate the actual regressors $X_j$.  Importantly, properties
\eqref{eq:orthogonality} are {\bf\em not} assumptions. They are consequences 
of the way in which these terms are defined. Their properties derive directly
from the decomposition described above and the fact that
$\Bbeta' \Xvec $ is the population OLS approximation to $Y$ and also
to $\mu(\Xvec)$.  This much holds in an assumption lean framework
without making any modeling assumptions whatsoever.

Because we assume an intercept to be part of the regressors ($X_0=1$), the
facts \eqref{eq:orthogonality} imply that all three terms are
marginally population centered:
\begin{equation} \label{eq:mean-zero}
  \E[\delta] ~=~ \E[\eta(\Xvec)] ~=~ \E[\varepsilon] = 0.
\end{equation}
However, it is {\bf\em not} true that $\E[\delta | \Xvec] = 0$, and
$\delta$ is {\bf\em not} independent of $\Xvec$ as would be the case
assuming a conventional error term in a linear model.  We have instead
$\E[\delta | \Xvec] = \eta(\Xvec)$, which, though marginally centered,
is a function of $\Xvec$ and hence not independent of the regressors
(unless it vanishes).  Similarly, although the noise $\varepsilon$ is
marginally centered and uncorrelated with the regressors, it is
generally dependent on the regressors, for example, in the form of
heteroskedasticity.

Concluding this section, we emphasize that the regressor variables 
have been treated as random and not as fixed.  The assumption 
lean framework has allowed a constructive decomposition 
that mimics some of the features of a linear model but replaces the 
usual assumptions made about ``error terms'' with orthogonality properties 
associated with the random regressors.  These properties are satisfied by 
the population residuals, the nonlinearity and the noise alike.  They are 
not assumptions. They are consequences of the decomposition.

\section{Random Regressors Interacting With a Nonlinear Response Surface} 
\label{sec:conspiracy}

Because in reality regressors are most often random variables that 
are as random as the response, it is a peculiarity of common statistical 
practice that such regressors are treated as fixed (Searle, 1970: Chapter~3).  
In probabilistic terms, this means that one conditions on the observed
regressors.  Under the frequentist paradigm, alternative datasets 
generated from the same model leave regressor values unchanged.
Only the response values change.  Consequently, regression 
models have nothing to say about the regressor distribution; 
they only model the conditional distribution of the response 
given the regressors, and there is no role for the regressor marginal
distributions.  This  alone might be seen by some as 
sufficient to justify conditioning on the regressors.  
There exists, however, a more formal justification, drawing on principles of
mathematical statistics: in any regression model, regressors are
ancillary for the parameters of the model, and hence, can be conditioned
on and treated as fixed.  This principle, however, has no validity
here because it applies only when the model is correct, which is precisely
the assumption discarded by an assumption lean framework.  Thus,
we are not constrained by statistical principles that apply only in a
model trusting framework.

\begin{figure}[htbp]
\begin{center}
   \includegraphics[scale=.29]{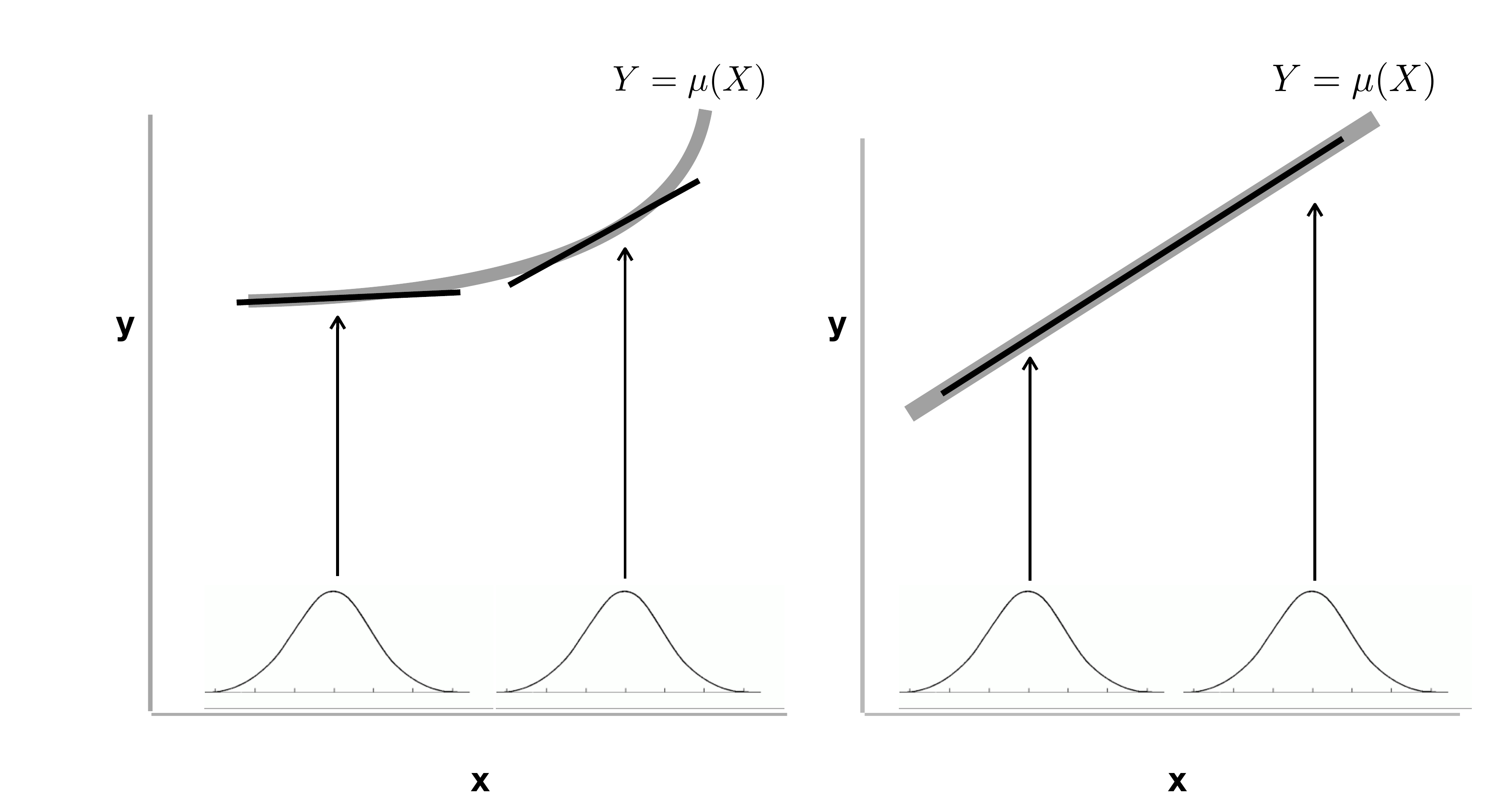} 
   \caption{Dependence of the Population Best Linear Approximation on the Marginal Distribution of the Regressors}
\label{fig:randomx}
\end{center}
\end{figure}

Ignoring the randomness of the regressors and their
marginal distribution is perilous under misspecification.
Figure~\ref{fig:randomx} shows why.  The left and right side pictures
both compare the effects of different regressor distributions for a
single regressor variable~$X$ in two situations: misspecification and
correct specification, respectively.  The left plot shows a case of
misspecification in which the true mean function $\mu(X)$ is nonlinear
and yet, a linear function is fitted.  The best linear approximation 
to the nonlinear mean function depends on the regressor
distribution $\PX$.  Therefore, the ``true parameters'' $\Bbeta$ ---
the slope and intercept of the best fitting line at the population ---
will also depend on the regressor distribution.  Specifically, one can
see that for the left marginal distribution the intercept is larger
and the slope is smaller than for the right marginal distribution.
This implies that under misspecification the regressor distribution
$\PX$, thought of as a ``non-parametric nuisance parameter,'' is no
longer ancillary.

The right side plot of Figure~\ref{fig:randomx} shows a case of
correct specification: The true mean function $\mu(X)$ (gray line) is
linear. Consequently, the best linear approximation is trivially the same
(black line) for both regressor distributions. In this case, the population 
marginal distribution of $X$ does not matter for the best linear approximation.  
There is one value for $\Bbeta$ no matter where the mass of $X$ falls.  This makes
the regressor distribution $\PX$ ancillary for the parameters
of the best linear fit.

The lessons from Figure~\ref{fig:randomx} generalize to multiple
linear regression with multivariate $\Xvec$,  but the effects
illustrated by the figure are magnified. Although one may argue that with a
single regressor it should be easy to diagnose the misspecification,
the challenges escalate for progressively larger numbers of
regressors, and become nearly impossible in ``modern'' settings where the
number of regressors exceeds the sample size, and data analysts tend to
gamble on sparsity.

Practical implications and questions arise immediately.  First, it is
the combination of a misspecified working model and random regressors
that produces the complications --- it matters where the regressors 
fall.  Second, one may wonder about the meaning of slopes
when the model is not assumed to be correct.  Third, what is the
use of predicted values $\hat{y} = \xvec' \Bbeta$?  Fourth, what form
should statistical inference take when there is no reliance on the
usual assumptions?  Specifically, how are standard errors and
associated $p$-values and confidence intervals affected by model
misspecification?  We will discuss possible answers to these questions
in the remaining sections.

\section{Conditional Mean Functions versus Regressor  Distributions}

The difficulties illustrated by Figure~\ref{fig:randomx} suggests
possibilities that may occur in various appplications, ranging
from modeling of grouped data to meta-analysis.  Consider the
following hypothetical scenarios that should serve as cautions when
interpreting models that are approximations.  In Section~\ref{sec:Interpretations} 
we will provide ways to interpret properly misspecified fitted linear functions.

Imagine a study of employed females and males in a certain
industry, with income as response and a scale measuring education
level as regressor.  Consider next the possibility that there is one
conditional mean function for income irrespective of gender,
but the mean function may be nonlinear in the education scale, as
illustrated by the left side picture in Figure~\ref{fig:randomx}.  A
data analyst may fit a linear model, perhaps because of convention,
 a high level of noise obscuring the nonlinearity, or a lack
of graphical data exploration.  The analyst may then find that
different slopes are required for males and females and may 
respond by including in the regression an interaction term between 
gender and education.  If, however, the truth is as stipulated, the usual 
interpretation of interaction effects would be misleading.  The 
driver of the gender difference is not in how income responds to 
education, but the education scale distribution by gender. Put in
different language, one may say that the real story is in the
consequences of an association between gender and education.

Imagine now meta-analysis of randomized clinical trials (RCTs). 
RCTs often produce different apparent treatment effects for the 
same intervention, sometimes called ``parameter heterogeneity.''  
Suppose the intervention is a subsidy for higher education, and the 
response is income at some defined end point. In two different
locales, the average education levels may differ. Consequently, 
in each setting the intervention works off different baselines. 
There can be an appearance of different treatment effects even 
though the nonlinear mean returns to education may be the same
 in the both locales.  The issue is, once again, that the difference 
 in effects on returns to education may not derive from different 
 conditional mean functions but from differences between 
 regressor distributions.

Apparent parameter heterogeneity also can materialize in the choice 
of covariates. The coefficient $\beta_1$ of the regressor $X_1$ is not 
to be interpreted  in isolation. $\beta_1$ depends on what other regressors
are included. In the simplest case, a regression on $X_1$ alone, differs
a regression on $X_1$ and $X_2$, Fitting a best approximation using 
$X_1$ alone produces a value of $\beta_1$ that would be the same 
in a regression on $X_1$ and $X_2$ {\em if} the two regressors were 
strictly uncorrelated.  In observational data, $X_1$ and $X_2$ are 
likely to be partially collinear or ``confounded'' to various degrees. It 
matters for $\beta_1$ whether $X_2$ is included in the regression.  
In the extreme, the coefficients $\beta_1$ obtained from the two 
regressions may have different signs, suggesting an instance of 
Simpson's paradox (see Berk et al.~2013, Section~2.1, for a more 
detailed discussion).  For our discussion, exclusion versus inclusion 
of $X_2$ can be interpreted as a difference in regressor distributions, 
namely, that of the marginal distribution of $X_1$ compared to the 
bivariate distribution of~$(X_1,X_2)$.

The prospect of more than one regressor can introduce
further complications in practice. If some of the candidate regressors 
are unrelated to the response, variable selection procedures 
are sometimes employed. A common and heroic assumption is 
that the mean function is correctly specified save for some 
unnecessary regressors, and that such regressors can be 
found and discarded. Perhaps the most fundamental 
difficulty is how to formulate a proper estimand for different 
models defined by different subsets of the regressors 
when none of the models is correctly specified. For us, that the target 
of estimation is the best approximation at the population for the
regressors selected. But within our approach, any approximation 
can be best by the least squares criterion, and the slopes of regressors
included in more than one approximation can vary dramatically. 
One must also properly handle the post selection inference. 

\section{Estimation and Standard Errors}

With i.i.d.~multivariate data $(Y_i,\Xvec_i)$ ($i=1,\ldots,n$) from
$\P$ on hand, one can apply OLS and obtain the plug-in estimate
$\Bbetahat = \Bbeta(\Phat)$ derived from \eqref{eq:population-OLS},
where $\Phat$ denotes the empirical distribution of the dataset.  By
multivariate central limit theorems, the regression estimates for the
slopes of the best linear approximation in the OLS sense are
asymptotically unbiased and normally distributed.  These estimates are
also asymptotically efficient in the sense of semi-parametric theory
(e.g, Levit 1976, p.~725, ex.~5; Tsiatis, 2006, p.~8 and ch.~4).

\subsection{Sandwich Standard Error Estimates}

The asymptotic variances of the OLS regression estimates in the
assumption lean i.i.d.~sampling framework deviate from those of linear
models theory which assumes linearity and homoskedasticity.  The
appropriate asymptotic variance is instead of a ``sandwich'' form
(White, 1980a):
\begin{equation} \label{eq:sand}
\AV[\Bbeta, \P] = \E[\Xvec \Xvec']^{-1} \;
                  \E[\delta^{2} \Xvec \Xvec'] \;
                  \E[\Xvec \Xvec']^{-1} .
\end{equation}
A plug-in estimator is obtained as follows:
\begin{equation} \label{eq:sand-est}
\AVhat = \AV[\Bbetahat, \Phat] = \left( \frac{1}{n} \sum_i \Xvec_i' \Xvec_i \right)^{-1} \;
                                 \left( \frac{1}{n} \sum_i r_i^2 \Xvec_i' \Xvec_i \right) \;
                                 \left( \frac{1}{n} \sum_i \Xvec_i' \Xvec_i \right)^{-1} ,
\end{equation}
where $r_i = Y_i - \Xvec_i' \Bbetahat$ are the sample residuals and
$\Phat$ is the empirical distribution of the data $(Y_i,\Xvec_i)$
$(i=1,...,n)$.  Equation~\eqref{eq:sand-est} is the simplest form of
sandwich estimator of asymptotic variance. More refined forms exist
but are outside the scope of this article.  Standard error estimates
for OLS regression coefficient estimates $\hat{\beta}_j$ are obtained
from \eqref{eq:sand-est} using the asymptotic variance estimate in the
$j$'th diagonal element:
\[
  S\!E_j = \left( \frac{1}{n} \AVhat_{\!j,j} \right)^{1/2} .
\]

A connection with linear models theory is as follows. If the truth is
linear and homoscedastic, hence the working model is correctly
specified to first and second order, then the sandwich
formula~\eqref{eq:sand} collapses to the conventional formula for
asymptotic variance due to
$\E[\delta^{2} \Xvec \Xvec'] = \sigma^2 \E[\Xvec \Xvec']$, which in
turn follows from
$\E[\delta^2|\Xvec] = \E[\epsilon^2|\Xvec] = \sigma^2$.  The result is
$\AV[\Bbeta, \P] = \sigma^2 \E[\Xvec \Xvec']^{-1}$.  This is the
``assumption laden'' form of asymptotic variance.

\subsection{Bootstrap Standard Error Estimates}

An alternative to standard error estimates based on the sandwich
formula is obtained from the nonparametric pairwise or $x$-$y$
bootstrap which resamples tuples $(Y_i,\Xvec_i)$.  It is transparently
assumption lean in that it relies for asymptotic correctness
essentially only on iid sampling of the tuples $(Y_i,\Xvec_i)$ and
some technical moment conditions.  The $x$-$y$ bootstrap therefore
applies to all manners of regressions, including~GLMs.

In stark contrast, the residual bootstrap is inappropriate. It necessarily 
assumes first order correctness, $\mu(\xvec) = \Bbeta' \xvec$, as well as
homoskedastic population residuals~$\delta$. The only step towards
assumption leanness is a relaxation of Gaussianity of the error
distribution.  Furthermore, it does not apply to other forms of
regression such as logistic regression.  The residual bootstrap
seems to be preferred by those who insist that one should condition 
on the regressors because they are ancillary. But as argued in
Section~\ref{sec:conspiracy}, the ancillarity argument assumes correct
specification of the linear regression model with independent errors,
controverting the idea that models are approximations rather than truths.

Sandwich and bootstrap estimators of standard error are identical in
the asymptotic limit, and on finite data they tend to be close.  Based
on either, one may perform conventional statistical tests and form
confidence intervals.  Although asymptotics are a justification for either,
one of the advantages of the bootstrap is that it lends itself to a
diagnostic to assessment of whether asymptotic normality is a reasonable
assumption for the analysis being undertaken. One simply creates normal 
quantile plots of bootstrap estimates obtained in the requisite simulations.  
Finally, bootstrap confidence intervals have be addressed in extensive research 
showing that there are variants demonstably higher order correct. See, for example, 
Hall (1992), Efron and Tibshirani (1994), Davison and Hinkley (1997).  An 
elaborate double-bootstrap procedure for regression is described in McCarthy et
al.~(2017).

\section{Interpretations}
\label{sec:Interpretations}

\subsection{Slopes from Best Approximations}

When the estimation target is the best linear approximation, one can
capitalize on desirable model-robust properties not available from
assumption laden, linear models theory.  The price is that
subject-matter interpretations address features of the best linear
approximation, not that of a ``generative truth'' --- which, as we have
emphasized, is often an unrealistic notion.\footnote{Even the
 minimal assumption of i.i.d.~sampling adopted here is often unrealistic.}

The most important interpretive issues are associated with the regression
coefficients of the best linear approximation.  The problem is that
the standard interpretation of a regression coefficient is not
strictly applicable anymore: It no longer holds that 
\begin{itemize}
\item[] {\em $\beta_j$ is
the average difference in $Y$ for a unit difference in $X_j$ at
constant levels of all other regressors~$X_k$.}
\end{itemize}
This statement uses the classical ``ceteris paribus'' (all things
being equal) clause, which only holds when the response function is
indeed linear.  One gets closer to a generalizable interpretation if
one modifies two details of this formulation as follows:
\begin{itemize}
\item[] {\em $\beta_j$ is the average ratio of differences in $Y$ over
  differences in $X_j$, linearly adjusted for all other
  regressors~$X_k$.}
\end{itemize}

To give this interpretation precise meaning, we focus on the
population target and consider for the moment a single regressor only.
The slope of the simple regression through the origin of $Y$ on $X$ is
then $\beta = \E[\, Y X \,] / \E[\, X^2 \,]$.  It is a matter of
elementary algebra to show that $\beta$ also equals the
following suggestive expression:
\begin{equation} \label{eq:weighted}
\beta = \E[\, w(X\!-\!X') \, \frac{Y\!-\!Y'}{X\!-\!X'}\,]
~~~~\textrm{where}~~~~  
w(X\!-\!X') = \frac{(X\!-\!X')^2}{\E[\, (X-\!X')^2 \,]} .
\end{equation}
Here $(X,Y)$ and $(X',Y')$ are two points drawn independently from the
joint $(X,Y)$ distribution, hence $(Y\!-\!Y')/(X\!-\!X')$ is the slope
of the line through this pair of points.  Formula \eqref{eq:weighted}
says that these pairwise slopes average out to the slope $\beta$ of
the OLS approximation when weighted proportionately to squared
horizontal distances $(X\!-\!X')^2$.  These weights lend more
influence to point pairs that are far from each other on the $X$-axis
and are, therefore, more informative for the slope.  Thus, $\beta$ is
indeed a weighted average of pairwise slopes.

To extend the interpretation to more than one regressor, we only need to
observe that the multiple regression coefficient $\beta_j$ is the
simple regression coefficient through the origin with regard to $X_j$,
linearly adjusted for all other regressors.  The same interpretation
is available for estimates $\hat{\beta}_j$ through its plug-in.  In either
case, formula \eqref{eq:weighted} provides an interpretation of slopes
as distance-weighted averages of pairwise slopes obtained from
linearly adjusted regressors (see Buja et al.~2016a for more details).

\subsection{Predicted Values $\hat{y}$ from Best Approximations}
\label{sec:prediction}

Also of interest are the predicted values at
specific locations $\xvec$ in regressor space, estimated as
$\hat{y}_{\xvec} = \hat{\Bbeta}' \xvec$.  In linear models theory,
for which the model is assumed correct, there is no bias if it is the
true response surface that is estimated by predicted values. That is,
$\E[\hat{y}_{\xvec}] = \Bbeta' \xvec = \mu(\xvec)$ because
$\E[\hat{\Bbeta}] = \Bbeta$, where $\E[\ldots]$ refers only to the
randomness of the response values $y_i$ with the regressor
vectors $\Xvec_i$ treated as fixed.

When the model is mean-misspecified such that 
$\mu(\xvec) \neq \Bbeta' \xvec$, then $\hat{y}_{\xvec}$ is an estimate
of the best linear approximation $\Bbeta' \xvec$, not $\mu(\xvec)$.
Hence, there exists bias $\mu(\xvec) - \Bbeta' \xvec = \eta(\xvec)$
that does not disappear with increasing sample size~$n$.  Insisting on
consistent prediction with linear equations at a specific location
$\xvec$ in regressor space is, therefore, impossible.

In order to give meaning to predicted values $\hat{y}_{\xvec}$ under
misspecification, it is necessary to focus on a population of future
observations $(Y_f,\Xvec_f)$ and to assume that it follows the same
joint distribution $\PYX$ as the earlier training data $(Y_i,\Xvec_i)$.
In particular, the future regressors are not fixed but random
according to $\Xvec_f \sim \PX$.  If this is a reasonable assumption,
then $\hat{y}_{\Xvec_f}$ is indeed the best linear prediction of
$\mu(\Xvec_f)$ and $Y_f$ for this future population under squared 
error loss.  Averaged over future regressor vectors, there is
no systematic bias because $\E[\eta(\Xvec_f)] = 0$ according to
\eqref{eq:mean-zero} of Section~\ref{sec:decomp}.\footnote{When
  regressors are treated as random, there exists a small estimation
  bias. $\E[\hat{\Bbeta}] \neq \Bbeta$ in general because
  $\E[(\frac{1}{n} \sum \Xvec_i' \Xvec_i)^{-1} (\frac{1}{n} \sum
  \Xvec_i Y_i) ] \neq \E[ \Xvec' \Xvec ]^{-1} \E[ \Xvec Y ] $, causing
  $\E[\hat{y}_{\xvec}] \neq \Bbeta' \xvec$ for fixed $\xvec$.
  However, this bias is of small order in~$n$ and shrinks rapidly
  with increasing~$n$.}  Asymptotically correct prediction intervals
for $Y_f$ do exist and, in fact, the usual intervals of the form
\begin{equation} \label{eq:pred}
P\!I_n(\xvec;K) ~:=~
\left[ \hat{y}_{\xvec} \pm K \cdot \hat{\sigma} \cdot \left(1 + \xvec'
  \left(\sum_{i=1\ldots n} \Xvec_i' \Xvec_i \right)^{-1} \xvec \right) \right]
\end{equation}
can be used.  However, the usual multiplier $K$ is based on linear
models theory with fixed regressors. It is not robust to
misspecification.  There exists a simple alternative for choosing $K$
that has asymptotically correct predictive coverage under
misspecification.  It can be obtained by calibrating the multiplier
$K_n$ empirically on the training sample such that the desired
fraction $1-\alpha$ of observations $(Y_i,\Xvec_i)$ falls in their
respective intervals.  One estimates $\hat{K}_n$ by satisfying an
approximate equality as follows, rounded to~$\pm 1/n$:
\[
\frac{1}{n} \cdot \#\left\{ i \in \{1,\ldots,n\} : ~Y_i \in P\!I(\Xvec_i;\hat{K}_n) \, \right\} ~\approx~ 1-\alpha .
\]
Under technical conditions, such multipliers will produce asymptotically correct
prediction coverage:
\[
\P\left[ \, Y_f \in P\!I(\Xvec_f;K_n) \, \right] ~\rightarrow~ 1-\alpha ~~~\textrm{as}~~~ n \rightarrow \infty ,
\]
where $\P[\ldots]$ accounts for randomness in the training data as
well as the future data.  (For more honest prediction, one might
consider a cross-validated version of calibration based on repeatedly
leaving out random portions of the data in fitting and estimating
$\hat{K}_n$ from those portions.)

This method of calibration is certainly not unique for the prediction
intervals of the form shown in \eqref{eq:pred}.  Its essential feature
is that the intervals form a one-parameter nested family:
~$P\!I_n(\xvec;K) \subset P\!I_n(\xvec;K')$~ for
$K \le K' \in \Reals$.  Thus, calibration for prediction can be applied
to many other forms of nested intervals, but those of \eqref{eq:pred}
are conventional and optimal under correct specification and yet, made
asymptotically accurate under misspecification by the simple device of
empirical calibration.

\subsection{Causality and Best Approximation}

Misspecification creates important challenges for causal
inference.  Consider first a randomized experiment with potential outcomes
$Y_1,Y_0$ for a binary treatment/intervention $C \in \{0,1\}$.  Because of
randomization, the potential outcomes are independent of the
intervention: $(Y_1,Y_0) \perp\!\!\!\perp C$.  Unbiased estimates of
the {\em Average Treatment Effect} (ATE) follow.  Pre-treatment
covariates $\Xvec$ can be used to increase precision (reduce standard
errors) only, similar to control variates in Monte Carlo (MC)
experiments.  It has been known for some time that the model
including the treatment $C$ and the pre-treatment covariates $\Xvec$
does not need to be correctly specified to provide correct estimation of
the ATE and (possibly) an asymptotic reduction of standard errors.  That
is, the model $Y \sim \tau C + \Bbeta' \Xvec$ may be arbitrarily
misspecified, and yet the ATE agrees with the treatment
coefficient~$\tau$.  (To yield a benefit, however, the covariates
$\Xvec$ must produce a useful increase in~$R^2$ 
or some other appropriate measure of fit, similar to
control variates in MC experiments.)

Now consider observational studies. There can be one or more variables
that are thought of as causal and which can at least in principle be
manipulated independently of the other covariates.  If there is just
one causal binary variable $C$, we are returned to a model of the form
$Y \sim \tau C + \Bbeta' \Xvec$, where it would be desirable for
$\tau$ to be interpretable as an average treatment effect (Angrist and
Pischke, 2009, Section 3.2).  These are always very strong claims that
often call for special scrutiny.  It is widely known that causal
inference properly can justified by assuming one of two sufficient conditions,
known as ``double robustness'' (see, e.g, Bang and
Robins 2005, Rotnitzki et al.~2012): (1)~Either the mean function for
$Y$ is correctly specified, which in practice means that there is no
``omitted variables'' problem for the response and that the functional
form of the conditional mean function for $Y$ is correct; or (2)~the
conditional probability of treatment (called the propensity score) can
be correctly modeled, which in practice means that there is no omitted
variables problem for treatment probabilities and that the (usually
logistic) functional form of the treatment probabilities is correct.
In either case, omitted variable concerns are substantively characterized and
not be satisfactorily addressed by formal statistical methods (Freedman, 2004).
There are certainly diagnostic proposals based on proxies for
potentially missing variables or based on instrumental variables, but 
the assumptions required a hardly lean. (see, for example, Hausman 1978).  
Misspecification of the functional form of the conditional response mean 
or the treatment probabilities are probably more appropriate for formal diagnostics.

In summary, causal inferences based observational data are fragile
because they depend on two things: (1)~correct model specification for
at least one of two response variables, the response mean or the treatment
probability, and (2)~correctly deciding which one it is.  Best
approximation under misspecification won't do.  As a consequence,
tremendous importance can fall to diagnostics of model~fit.  See Buja et
al.~(2016b) for some useful diagnostics that are applicable in all
types of regressions for i.i.d.~observations. But then, we are no
longer doing assumption lean regression.  

\section{Generalizations}

Focussing regression functionals provides possibilities for
generalization.  A first and readily apparent generalization is to statistical
functionals other than slopes of the best linear approximation.
Conditional variances come to mind. For example, if medical expenses 
are made a function of income, there can be more variation around the 
conditional mean for high-income households because they can afford 
more discretionary medical procedures.

A second and equally apparent generalization is to regressions other
than linear OLS, such as generalized linear models and in particular,
to linear logistic regression. The response is now binary, which suggests 
modeling conditional Bernoulli probabilities with a suitable link function
and a cost function other than least squares.  Interpreting the parameters 
as functionals allows the conditional distributions of the binary response 
to be largely arbitrary. One need not assume the logistic model is correct.  
The working model  becomes a heuristic that produces a plausible cost function.

In detail, for a binary response $Y \in \{0,1\}$ one models the logit
of the conditional probabilities
\[
p(\xvec) ~=~ \P[Y \!=\! 1|\Xvec \!=\! \xvec] ~=~ \E[Y|\Xvec \!=\! \xvec] 
\]
with a linear function of the regressors:
\[
\logit(p(\xvec)) = \log(p(\xvec)/(1-p(\xvec))) ~\approx~ \Bbeta' \xvec .
\]
The ${\rm logit}(p(\xvec))$ maps in reverse to a model of the conditional probabilities
via the sigmoid function that is the natural link function of
the Bernoulli model:
\[
p(\xvec) ~\approx~ \phi(\Bbeta' \xvec), ~~\textrm{where}~~ \phi(t) = \exp(t) / (1+\exp(t)) .
\]
In both cases, we use ``$\approx$'' rather than ``$=$'' to indicate
the use of an approximation that allows varying degrees of
misspecification.

The negative log-likelihood of the model when $n \rightarrow \infty$
results in a population cost function whose minimization produces the
statistical functional (= estimand = ``population parameter'') as
follows:
\begin{equation} \label{eq:logistic}
\Bbeta(\P) ~:=~ \argmin_{\Bbeta \in \Reals^{p+1}}
  \E \left[ \log\left( 1 + \exp\big(\Xvec'\Bbeta\big) \right) - \big(\Xvec'\Bbeta\big) \, Y \right] .
\end{equation}
The usual estimates $\hat{\Bbeta}$ are obtained by plug-in,
replacing the expectation with the mean over the observations and
thereby returning to the negative log-likelihood of the sample.

Interpretations and practice follow much as earlier, with the added
complications caused by the nonlinear link function $\phi(t)$. The
estimate $\hat{\Bbeta}$ correspond to a best approximiation
$\phi(\Bbeta'\xvec)$ of the true response surface $p(\xvec)$.  The
estimates $\hat{p}(\xvec) = \phi(\hat{\Bbeta}'\xvec)$ are able to
target only the best approximation $\phi(\Bbeta'\xvec)$ where
$\Bbeta = \Bbeta(\P)$, not the true $p(\xvec)$.  The approximation
discrepancy $p(\xvec) - \phi(\Bbeta'\xvec)$ does not vanish with more
data, $n \rightarrow \infty$.

For standard errors, and statistical tests and confidence intervals
based on them, one should use the appropriate sandwich estimators or
standard error estimates obtained from the nonparametric $x$-$y$
bootstrap.  Both have asymptotic justification under i.i.d.~sampling
of the tuples $(Y_i,\Xvec_i)$ for the best approximation
target~$\Bbeta(\P)$.

Finally, under misspecification the regression functional $\Bbeta(\P)$
will generally, as before, depend on the distribution regressor $\PX$.  The
regressors cannot be treated as ancillary and not held fixed.  The
regression functional $\Bbeta(\P)$ can have different values depending
on where in the regressor space the data fall.

\section{An Empirical Example Using Poisson Regression}

In order to show the ease with which the functional view can be
applied within the generalized linear model, we consider an
application using Poisson regression.  The estimand is a best
log-linear approximation of the true response surface, not the true
response surface itself.

The particular Poisson regression will be applied to data from a
criminal justice agency.  Here is some context. An important feature
of an arrest is the charges that police choose to file.  One crime
event can lead to one charge or many.  Each charge for which there is
a guilty plea or guilty verdict will have sanctions specified by
statute.  For example, an aggravated robbery is defined by the use of
a deadly weapon, or an object that appears to be a deadly weapon, to
take property of value.  If that weapon is a firearm, there can then
be a charge of aggravated robbery and a second charge of illegal use
of a firearm.  There can be penalties for each. A greater number of
charges can be used by prosecutors as plea bargaining chips and can
place an offender at greater risk of sanction.  In this illustration,
we consider correlates of the number of charges against an offender
filed by the police.\footnote { Although the charges are specified by
  the police, they are typically reviewed by prosecutors who may
  change the charges.  }
 
The dataset for our illustration contains 10,000 offenders arrested
between 2007-2015 in a particular urban jurisdiction.  The data are a
random sample from over three hundred thousand offenders arrested in
the jurisdiction during those years.  This pool is sufficiently large
to make an assumed infinite population and iid sampling good
approximations.  During that period, the governing statutes,
administrative procedures, and mix of offenders were effectively
unchanged -- there is a form of criminal justice stationarity.  We use
as the response variable the number of charges associated with the
most recent arrest.  Several regressors are available, all thought to
be related to the outcome.  Many other relevant regressors are not
available (e.g., the consequences of the crime for its victims).

The approximation adopted here is a Poisson regression. It is a working
model about which we make no claim that it is correctly specified.  Consequently,
we forfeit any causal claims, and we are not proposing any of the
regressors as manipulable interventions.  Finally, our assumption lean
framework does not require that the response follows a conditional
Poisson distribution.  One reason for violating Poissonness is that
the binary events constituting the counts do not need to be
independent.  Indeed, independence would be unrealistic. If the crime
is an armed robbery, for example, the offender would be charged with
aggravated robbery and a weapons offense.  Ordinarily, such dependence
would be a concern.

\medskip

The results of the Poisson regression are shown in
Table~\ref{tab:poisson}.  The columns contain, from left to right, the
following quantities:
\begin{enumerate} \itemsep -0.2em
\item
the name of the regressor variable, 
\item
the usual Poisson regression coefficient,
\item
the conventional standard errors,
\item
the associated p-values, 
\item
standard errors computed using a nonparametric $x$-$y$ bootstrap,
\item
standard errors computed with the sandwich estimator, and
\item
the associated sandwich p-values.
\end{enumerate}

\begin{table}
\small
\begin{center}
\begin{tabular}{lcccccc}
\hline
& Coeff & SE & p-value & Boot.SE & Sand.SE & Sand-p \\
 \hline
(Intercept) & 1.8802 & 0.0205 & 0.0000 & 0.0522 & 0.0526 & 0.0000 \\
 Age & -0.0147 & 0.0006 & 0.0000 & 0.0016 & 0.0016 & 0.0000 \\
 Male & 0.0823 & 0.0127 & 0.0000 & 0.0284 & 0.0299 & 0.0058 \\
 Number of Priors & 0.0031 & 0.0002 & 0.0000 & 0.0005 & 0.0005 & 0.0000 \\
 Number of Prior Sentences & 0.0002 & 0.0016 & 0.8868 & 0.0040 & 0.0039 & 0.9519 \\
 Number of Drug Priors & -0.0138 & 0.0008 & 0.0000 & 0.0021 & 0.0020 & 0.0000 \\
 Age At First Charge & 0.0028 & 0.0009 & 0.0012 & 0.0022 & 0.0021 & 0.1935 \\
\hline
\end{tabular}
\caption{Poisson Regression Results for The Number of Crime Charges (n=10,000)}
\label{tab:poisson}
\end{center}
\end{table}

Even though the model is likely misspecified by conventional standards
for any number of reasons, the coefficient estimates for the
population approximation are asymptotically unbiased for the
population best approximation.  In addition, asymptotic normality
holds and can be leveraged to produce approximate confidence intervals
and p-values based on sandwich or $x$-$y$ bootstrap estimators of
standard error.

With 10,000 observations, the asymptotic results effectively apply.
None of this would be true for inferences based on assumption-laden
theories that assume the working model to be correct.

The marginal distribution of the response (number of charges) is
skewed to the right: The mean number of charges range from 1 to 40,
with a mean of 4.7, a standard deviation of 5.5.  Most offenders have
a relatively small number of charges, but a few offenders have many.

Table~\ref{tab:poisson} shows that some of the bootstrap and sandwich
standard errors are rather different from the conventional standard
errors, indicating indirectly that the conditional Poisson model is
misspecified (Buja et al. 2016a).  Moreover, there is a reversal of
the test's conclusion for ``Age at First Charge'' (i.e., the earliest
arrest that led to a charge as an adult). The null hypothesis is
rejected with conventional standard errors but is not rejected with a
bootstrap or sandwich standard error.

This correction is sensible because a natural expectation would
have been that the slope of ``Age At First Charge'' should be
negative, not positive. Typically individuals who have an early arrest
and charge are more likely to commit crimes later on for which there
can be multiple charges.

In the present Poisson model, exponentiated regression coefficients
are multipliers of the charge counts.  We interpret each regressor
accordingly if the null hypothesis is rejected based on
sandwich/bootstrap standard errors:
\begin{itemize}
\item {Age}: Starting at the top of Table~\ref{tab:poisson}, each additional
  year of age multiplies the average charge count by a factor of 0.98.
  Ten additional years of age reduces the count by a factor of 0.86.
  Older offenders have fewer charges perhaps because the crimes they
  tend to commit are different from the crimes younger offenders
  commit.  For instance, the crimes of younger offenders may be more
  likely to be gang-related.
\item {Male}: Men on the average have a greater number of charges than
  women.  The multiplier is 1.08, which means that there is about a
  8\% average difference.  Compared to a women at the women's mean of
  4.7 charges, a man would on the average have 5.1 charges, holding
  all other covariates constant.
\item {Number of Priors}: To get the same 8\% increase from the
  exponentiated regression coefficients for the number of all prior
  arrests takes an increment of about 25 priors.  Such increments are
  common.  About 25\% of the offenders in the data are first offenders
  (i.e., no prior arrests), and another 30\% have 25 or more prior
  arrests.  A gap of 25 priors is common in these data.
\item {Number of Drug Priors}: Offenders with a greater number of
  prior arrests for drug offenses on the average have fewer charges
  after controlling for the other covariates.  Drug offenders often
  have a large number of such arrests, so the small coefficient of
  -0.0138 matters.  For 20 additional prior drug arrests, the average
  charge count is multiplied by a factor of .76.  A long history of
  drug abuse can be debilitating so that the crimes committed are far
  less likely to involve violence and often entail little more than
  drug possession.
\end{itemize}
In summary, offenders who are young males with many prior arrests not
for drug possession, will on average have substantially more criminal
charges.  Such offenders perhaps are likely to be disproportionately
arrested for crimes of violence in which other felonies are committed
as well.  Therefore, a larger number of charges would be expected.

What about causal inference?  It makes little sense to envision
manipulating any of the regressors with all other regressors fixed.
The different measures of prior record and age at first arrest are in
the past.  Even if one could alter them, many other upstream
regressors would be altered as well.  Gender and age are inherent
features of an offender.  It also makes little sense to assume that
all excluded regressors are independent of those that are included.
For example, there is no measure of gang membership in the data.  But
even if gang membership were a regressor, one surely would not have
exhausted the list of potential confounders.  

If causal inference is off the table, what have we gained?  For one
thing, we acquired a general understanding of certain associations
among variables of interest.  In particular, it is instructive that
the findings are largely consistent with past research.  Note that we
still care about the signs of the regression coefficients as
indicators of the direction of association between the response and a
regressor adjusted for the presence of the other regressors.  What has
changed is that the rigidity of interpretation based on correct model
specification is abandoned and replaced by greater realism and
skepticism about what a regression coefficient is able to convey.

As for predictive use of the regression, the results could in principle
be applied in real settings to inform risk assessments in future cases.
We qualify such use with the cautions of Section~\ref{sec:prediction}
according to which prediction covergance is for a population of future
cases that behave like the past population, not any fixed location in
regressor space.  To this end, prediction intervals should be directly
calibrated on the data, not based on model trusting theory.  The
widths of such intervals can inform users whether the model contains
any actionable information at all.

We have seen indirect indications of model misspecification
Traditional model-trusting standard errors differ from assumption lean
sandwich and bootstrap standard errors.  Because of model
misspecification, it is likely that the parameters of the best fitting
model depend on where in regressor space the mass of the
$X$-distribution falls.  This raises concerns about the performance of
out-of-sample prediction.  If the out-of-sample data are not derived
from a source stochastically similar to that in the analyzed sample,
then these predictions may be wildly inaccurate.

\section{Conclusions}

Treating models as best approximations should replace treating models as 
if there are correct.  Best approximations proceed with a fixed model explicitly
acknowledging approximation discrepancies, sometimes called ``model bias,'' 
which do not disappear with more data.  The model bias, however, does not 
create an asymptotic bias in estimates of best approximations.  Parameters of best 
approximations are estimated with bias that disappears at the usual 
rapid rate.

In regression, a fundamental feature of best approximations
is that they depend on regressor distributions.  It follows that 
one cannot condition on regressors and treat regressors as fixed.  
Regressor variability must be included in treatments of the sampling 
variability for any estimates. This can be achieved by using model robust 
standard error estimates in statistical tests and confidence intervals.  
Two choices are readily available: sandwich estimators and bootstrap-based 
estimators of standard errors.  For the latter, a strong arguments favor
the nonparametric $x$-$y$ bootstrap over the residual bootstrap, 
because conditioning on the regressors and treating them as fixed
 is incorrect when there is misspecification.

In this article, we also offered at least three ways in which best
approximations can be informative in practice: (1)~model
parameters are re-interpreted as regression functionals,
(2)~predictions are for populations rather than at fixed
regressor locations, and (3)~there exist well-known connections 
between misspecification with causal inference.

In summary, it is easy to agree with G.E.P.~Box' famous dictum, 
but there are consequences affecting the mechanics
of statistical inference and the interpretations of statistical
estimates. Assume and proceed statistics does not suffice.
Nor does hand waving. 

\section*{References}
\begin{description}
\item
Angrist, J.D. and Pischke. J.-S. (2009). \textit{Mostly Harmless Econometrics: An Empiricist's Companion}. Princeton University Press.
\item
Benjamin, D. J., Berger, J. O.,  Johannesson, M., Nosek, B. A., Wagenmakers, E.--J., Berk, R., Bollen, K. A., Brembs, B., Brown, L., Camerer, C., Cesarini, D., Chambers, C. D., Clyde, M., Cook, T. D., De Boeck, P., Dienes, Z., Dreber, A., Easwaran, K., Efferson, C., Fehr, E., Fidler, F., Field, A. P., Forster, M., George, E. I., Gonzalez, R., Goodman, S., Green, E., Green, D. P., Greenwald, A., Hadfield, J. D., Hedges, L. V., Held, L., Ho, T.--H., Hoijtink, H., Jones, J. H., Hruschka, D. J., Imai, K., Imbens, G., Ioannidis, J. P. A., Jeon, M., Kirchler, M., Laibson, D., List, J., Little, R., Lupia, A., Machery, E., Maxwell, S. E., McCarthy, M., Moore, D., Morgan, S. L., Munafó, M., Nakagawa, S., Nyhan, B., Parker, T. H., Pericchi, L., Perugini, M., Rouder, J., Rousseau, J., Savalei, V., Schönbrodt, F. D., Sellke, T., Sinclair, B., Tingley, D., Van Zandt, T., Vazire, S., Watts, D. J., Winship, C., Wolpert, R. L., Xie, Y., Young, C., Zinman, J., and Johnson, V. E. (2017, in press). ``Redefine Statistical Significance.''  \textit{Nature Human Behavior}. 
\item Bang, H., and Robins, J.M. (2005) 
``Doubly Robust Estimation in Missing Data and Causal Inference,'' 
\textit{Biometrics} 61, 962--972.
\item
 Bachoc, F., Preinerstorfer, D., and Steinberger, L. (2017).
``Uniformaly valid confidence intervals post-model-selection.''
arXiv:1611.01043.
\item 
Berk, R.A., Brown, L., Buja, A., Zhang, K., and Zhao, L. (2013).
``Valid Post-Selection Inference.''
\textit{The Annals of Statistics} 41(2): 802--837.
\item
Berk, R.A. (2003) \textit{Regression Analysis: A Constructive Critique}. Newbury Park,  CA.: Sage.
\item
Berk, R.A., Sorenson, S.B., and Barnes, G. (2016). ``Forecasting Domestic Violence: A Machine Learning Approach to Help Inform Arraignment Decisions. \textit{Journal of Empirical legal Studies} 31(1): 94--115.
\item
Box, G.E.P.  (1976). ``Science and Statistics.'' \textit{Journal of the American Statistical Association} 71(356): 791--799.
\item
Buja, A., Berk, R., Brown, L., George, E., Pitkin, E., Traskin, M., Zhan, K., and Zhao, L. (2016a). ``Models as Approximations --- Part I: A Conspiracy of Nonlinearity and Random Regressors in Linear Regression.'' arXiv:1404.1578
\item
Buja, A., Berk, R., Brown, L., George, E., Arun Kuman Kuchibhotla, and Zhao, L. (2016b). ``Models as Approximations --- Part II: A General Theory of Model-Robust Regression.'' arXiv:1612.03257
\item
Davison, A.C., and Hinkley, D.V. (1997).
\textit{Bootstrap Methods and Their Application},
Cambridge University Press.
\item
Cox, D.R. (1995). ``Discussion of Chatfield'' (1995). \textit{Journal of the Royal Statistical Society}, Series A 158 (3), 455--456.
\item 
Efron, B., and Tibshirani, R.J. (1994).
\textit{An Introduction to the Bootstrap},
Boca Raton, FL: CRC Press.
\item
Fisher, R.A. (1924). ``The Distribution of the Partial correlation Coefficient.'' \textit{Metron} 3: 329-332.
\item
Freedman, D.A. (1981). ``Bootstrapping Regression Models.'' \textit{Annals of Statistics} 9(6): 1218--1228.
\item
Freedman, D.A. (2004). ``Graphical Models for Causation and the Identification Problem.'' \textit{Evaluation Review} 28: 267--293.
\item
Freedman, D.A. (2009). \textit{Statistical Models} Cambridge, UK: Cambridge University Press.
\item 
Hall, P. (1992).
\textit{The Bootstrap and Edgeworth Expansion.} (Springer Series in Statistics) New York, NY: Springer Verlag.
\item Hausman, J. A. (1978). 
``Specification Tests in Econometrics.'' \textit{Econometrica} 46 (6): 1251-–1271. 
\item 
Hong, L., Kuffner, T.A., and Martin R. (2016).``On Overfitting and Post-Selection Uncertainty Assessments.'' \textit{Biometrika} 103: 1--4.
\item
Imbens, G.W., and Rubin, D.B., (2015).  \textit{Causal Inference Statistics, Social, and Biomedical Sciences: An Introduction}. Cambridge: Cambridge University Press.
\item
Koenker, R. (2005). \textit{Quantile Regression} Cambridge: Cambridge University Press. 
\item
  Kuchibhotla, A.K., Brown L.D., Buja, A., George, E., Zhao, L. (2018).
  ``A Model Free Perspective for Linear Regression: Uniform-in-model Bounds for Post Selection Inference.''
  arXiv:1802.05801
\item
Leamer, E.E. (1978). \textit{Specification Searches: Ad Hoc Inference with Non-Experimental Data}. New York, John Wiley.
\item
  Levit, B. Y. (1976). 
  ``On the efficiency of a class of non-parametric estimates.''
  Theory of Probability \& Its Applications, 20(4):723--740.
\item
McCarthy, D., Zhang, K., Berk, R.A., Brown, L., Buja, A., George, E., and Zhao, L.(2017). ``Calibrated Percentile Double Bootstrap for Robust Linear Regression Inference.'' \textit{Statistica Sinica}, forthcoming
\item Rotnitzky, A., Lei, Q., Sued, M. and Robins, J. M. (2012). 
``Improved Double-Robust Estimation in Missing Data and Causal Inference Models,''
\textit{Biometrika} 99, 439–-456.
\item
Rubin, D. B. (1986). ``Which Ifs Have Causal Answers.'' \textit{Journal of the American Statistical Association} 81: 961--962.
\item
Searle, S.R. (1970). \textit{Linear Models}. New York: John Wiley.
\item
Lee,~J.~D., Sun,~D.L., Sun,~Y., and Taylor,~J.E. (2016).
``Exact Post-Selection Inference, with Application to the Lasso.''
\textit{The Annals of Statistics} 44(3): 907--927.
\item
Tsiatis, A.A. (2006). \textit{Semiparametric Theory and Missing Data}. New York: Springer.
\item
Wager, S., and Athey, S. (2017). ``Estimation and Inference of Heterogeneous Treatment Effects using Random Forests.'' \textit{Journal of the American Statistical Association}, in press.
\item
White, H. (1980a). ``Using Least Squares to Approximate Unknown Regression Functions.'' \textit{International Economic Review} 21(1): 149--170. 
\item
White H. (1980b). ``A Heteroskedasticity-Consistent Covariance Matrix and a Direct Test for Heteroskedasticity.'' \textit{Econometrica}, 48, 817--838.

\end{description}
\end{document}